\documentclass[aps,prl,twocolumn,superscriptaddress,showpacs,amsmath,amssymb,longbibliography]{revtex4-2}
\usepackage{graphicx}
\usepackage[colorlinks=true,citecolor=blue,linkcolor=blue,urlcolor=blue,bookmarks=true]{hyperref}

\usepackage[utf8]{inputenc}
\usepackage[english]{babel}
\usepackage{amsmath}
\usepackage{xcolor}
\usepackage{soul}

\definecolor{nred} {RGB}{224,0,0}
\definecolor{nblue} {RGB}{28,130,185}
\definecolor{dgreen}{RGB}{78,138,21}
\definecolor{norange}{RGB}{230,120,20}

\newcommand{\ch}[1]{\textcolor{black}{#1}}
\newcommand{\zm}[1]{\textcolor{black}{#1}}

\begin{document}

\title{Thouless approach in transport in integrable and perturbed easy-axis Heisenberg chains}  
\author{J. Paw\l{o}wski}
\affiliation{Institute of Theoretical Physics, Faculty of Fundamental Problems of Technology, Wroc\l{a}w University of Science and Technology, 50-370 Wroc\l{a}w, Poland}
\author{M. Mierzejewski}
\affiliation{Institute of Theoretical Physics, Faculty of Fundamental Problems of Technology, Wroc\l{a}w University of Science and Technology, 50-370 Wroc\l{a}w, Poland}
\author{P. Prelov\v{s}ek}
\affiliation{Jo\v{z}ef Stefan Institute, SI-1000 Ljubljana, Slovenia}

\begin{abstract}
We study transport in spin chains employing the Thouless approach based on the level sensitivity to the boundary conditions, $R$. Although  spin transport in the integrable easy-axis XXZ model is diffusive, corresponding $R$ is much closer to ballistic chains than to chaotic diffusive systems.  In the case of the grand canonical ensemble, this observation can be rigorously justified, while in the case of the canonical ensemble it can be demonstrated by numerical calculations.
Integrability breaking perturbation (IBP) strongly
reduces $R$ which reveals a pronounced minimum at the crossover  from anomalous diffusive to normal dissipative transport. This minimum coincides with the onset of the universality of the  random matrix theory. 
Results for various IBP  suggest a discontinuous jump  of the spin conductivity in the thermodynamic limit, and moreover that its value \ch{is proportional to the strength of the IBP}.  
     
\end{abstract}

\maketitle

\noindent {\it Introduction.}  
The 
properties of integrable quantum lattice models, the prominent example
being the one-dimensional, anisotropic Heisenberg XXZ spin-$1/2$ chain, have in last decades reached
higher level of understanding  (for recent review see e.g. \cite{bertini21}) through the 
extended concepts of local \cite{zotos97} and quasilocal conserved quantities (CQ) 
\cite{prosen11,prosen13}, via the analytical results emerging from  the generalized hydrodynamics (GHD)
\cite{bertini16,castro-alvaredo16,ilievski17,bulchandani18,denardis18} and with the application of  
various powerful numerical methods \cite{zotos96,znidaric11,mierzejewski11,karrasch14,steinigeweg15,prelovsek22}.
While the standard argument is that mastering integrable models will be the crucial
step to the proper description of closely related perturbed models, the advance
in this direction is so far modest. In the case of the  Heisenberg-type spin 
chains it appears evident that the integrability-breaking perturbations (IBP) suppress 
the finite-temperature ballistic transport, characterized by the spin stiffness $D(T>0) >0$,
turning it into the normal dissipative one  \cite{castella96,jung06,jung07,znidaric20,mierzejewski22}. Still,  analytical approaches and tools dealing with such perturbed systems are so far quite restricted \cite{friedman20,bastianello21,denardis21} .

The anisotropic Heisenberg chain in the easy-axis \mbox{($\Delta >1$)}
regime represents a quite different challenge, with vanishing spin conductivity (and diffusion) at temperature $T \to 0$, as well as vanishing stiffness $D(T>0) = 0$, but revealing finite
spin diffusion constant ${\cal D}$ even at high $T$ \cite{znidaric11,steinigeweg12,karrasch14,gopalakrishnan19}. The diffusion
is still anomalous in origin \cite{prelovsek04,znidaric14, medenjak17},
and the transport is dissipationless \cite{prelovsek22}.  Taking into account  open questions in this problem, the introduction of IBP leads to even wider range of scenarios. While there appears general consensus that in large enough systems (system size $L \to \infty$)
the spin transport  becomes normal (i.e. dissipative), its variation with IBP strength $g$ remains controversial. While some  recent results indicate on the continuous variation of ${\cal D}$ with $g$ \cite{kraft24},
there is also evidence for its discontinuity at $g \to 0 $ \cite{prelovsek22,denardis22}. Moreover,
introducing in the system a field $F >0$ \cite{mierzejewski11} even favors the vanishing  of the spin conductivity
with the vanishing $F \to 0$.

In this Letter we use an alternative approach to the spin transport 
both in integrable and perturbed  XXZ chains at $\Delta >1$. Since these spin chains are equivalent to the models of interacting fermions, 
 one can introduce a nonzero flux \cite{kohn64} that probes also system's response to changing of the boundary conditions. In particular, we employ the Thouless level sensitivity $R$. It was originally introduced into single-particle models to distinguish between 
conductors with $R \gg 1$ and insulators (localized systems)  with $R \to 0$  \cite{edwards72} and has recently been applied also to interacting disordered systems
\cite{prelovsek23}. 

The level sensitivity distinguishes between ballistic and normal diffusive transport. In the former case it grows (up to logarithmic corrections)  as the inverse of the mean level spacing,  $R \propto 1/\Delta \epsilon$ whereas in generic diffusive systems this dependence is much weaker, namely $R\propto 1/\sqrt{\Delta \epsilon}$. We show that the diffusive transport in the easy-axis integrable XXZ chain ($\Delta > 1$) is anomalous in this respect.
Namely, diffusive integrable chains reveal \mbox{$R \propto 1/\Delta \epsilon$}   in the grandcanonical ensemble (GCE),  as well as in the  canonical ensemble (CE)  at  total $S^z_{tot} =0$. Therefore, the level sensitivity of diffusive integrable system differs from $R$ in the ballistic systems only by logarithmic corrections, $\log(\Delta \epsilon)$, and is much larger than in generic diffusive models. 
Large values of the level sensitivity in GCE, \mbox{$R_{\text{GCE}} 
\gg 1$}, are explained by strict lower bound.

We also observe that level sensitivity \ch{ in CE, $R_{\text{CE}}$, obtained for} finite systems exhibits a nonmonotonic dependence on the strength of IBP, $g$. The  
minimum at $g=g^{*}(L)$ coincides with the crossover where the finite-size system starts to comply with the random matrix theory (RMT) \cite{brody81,wilkinson90}. Therefore, the minimum of $R$ allows us to delimit two regimes: $g>g^{*}(L)$ when the system behaves as normal dissipative and $g<g^{*}(L)$ when the system properties are still dominated by finite-size effects.  
Taking into account the restriction, $g>g^{*}(L)$,
one can then follow in a controlled way the variation with $g$ of the 
d.c.~spin conductivity in CE, $\sigma_{\text{CE}}$. Our results for two  classes of IBP seem to support a universal limit,
i.e., vanishing of $\sigma_{\text{CE}}( g \to 0)$ in the thermodynamic limit. 

\noindent {\it XXZ Model with flux.} 
We consider the XXZ spin \mbox{($s=1/2$)} chain of length $L$
with periodic boundary conditions (PBC),  introducing also 
a flux, $\varphi$, 
\begin{equation}
H = J \sum_l \left[ \frac{1}{2}( \mathrm{e}^{-i \varphi} S^+_{l+1} S^-_l + \mathrm{H.c.}) + \Delta S^z_{l+1} S^z_l  \right]
+ g H^\prime. \label{xxz}
\end{equation}
 We consider the  easy-axis regime with 
 $\Delta >1$ and  set $J=1$.
The spin model can be mapped on the model of
interacting spinless fermions, where $\varphi L$ represents a magnetic flux  through the ring \cite{kohn64,edwards72,prelovsek22}. We analyze two 
forms  of IBP. In the main text, we discuss the next-nearest neighbor (nnn) interaction with $g=\Delta_2$ and  $H^\prime = \sum_l S^z_{l+2} S^z_l$, whereas in the Supplemental Material \cite{supmat} we study  the nnn exchange with $g=J_2$ and $H^\prime = \sum_l  [\mathrm{e}^{-2 i \varphi} S^+_{l+1} S^z_l S^-_{l-1} +  \mathrm{H.c.}]$  which maps to a nnn hopping in the fermionic chain. 

\begin{figure}[h]
  \centering
  \includegraphics[width=\linewidth]{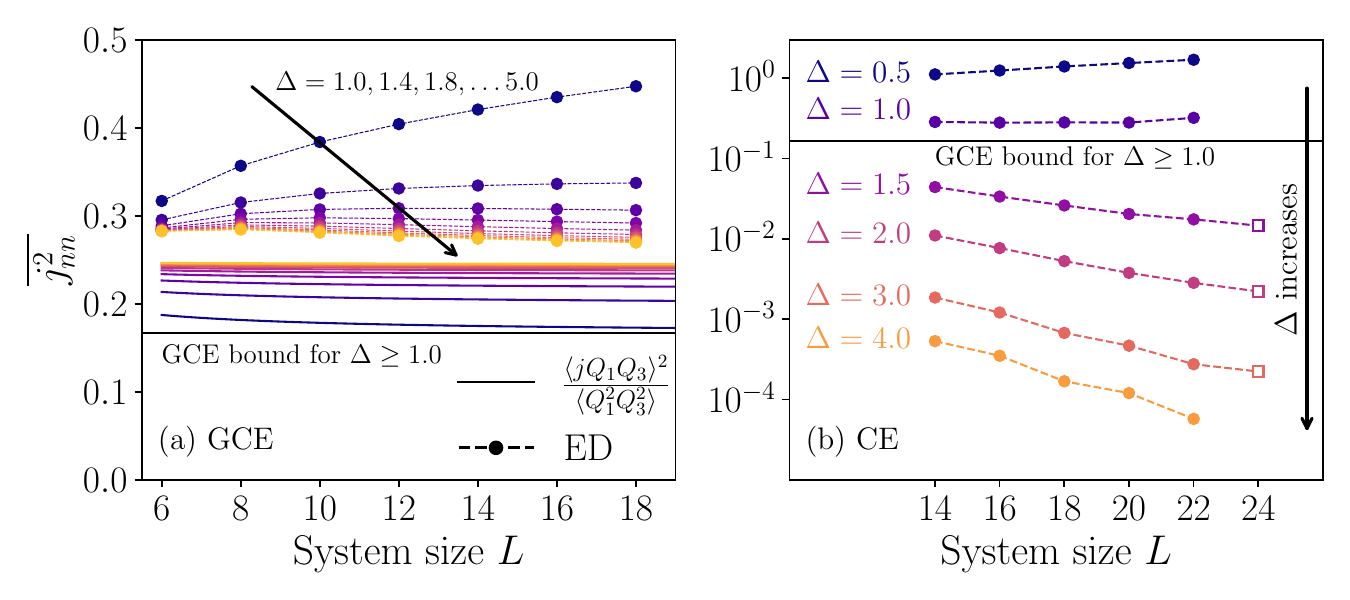}
  \caption{(a) GCE:  bound  from Eq.~\eqref{bound} (lines), compared with ED results (points) for  the diagonal matrix elements, $\overline{j^2_{nn}}$. (b) CE:  $\overline{j^2_{nn}}$ obtained via ED for the $S^z_{tot}=0$ and various $\Delta = 0.5 - 4$. The horizontal line is the GCE bound,  Eq.~\eqref{bound},  that holds for $\Delta \ge 1$ and any $L$.}
  \label{fig1}
\end{figure}

\noindent{\it Level sensitivity and spin transport.}  
The gauge  transformation $S^{\pm}_l\to e^{\pm i l \varphi} S^{\pm}_l$, $S^z_l\to S^z_l$ applied to the Hamiltonian~\eqref{xxz} eliminates the flux dependence on all sites except the boundary term that takes the form
\mbox{$J/2(S^+_1S^-_Le^{-i \varphi L}+{\rm H.c.})$}. Then, the energy spectra, \mbox{$H|n\rangle=\epsilon_n(\varphi) |n\rangle$}, obtained for $\varphi=0$ and
$\varphi=\delta \varphi=\pi/L$ correspond, respectively, to periodic and antiperiodic ($J\to- J$) boundary conditions. Consequently, the level sensitivity to the boundary conditions is related to the derivative $\epsilon'_n(\varphi)$ that can be further simplified using an identity for the spin-current operator, $j = {\rm d} H/ {\rm d} \varphi$. Following the reasoning from Refs. \cite{kohn64, castella95} one finds
\begin{equation}
\epsilon'_n(\varphi)= \frac{{\rm d} }{{\rm d} \varphi}
\langle n | H |n\rangle=\langle n | j |n\rangle+\epsilon_n \frac{{\rm d} }{{\rm d} \varphi} \langle n |n\rangle=\langle n | j |n\rangle.
\end{equation}
The level sensitivity introduced by Thouless in Ref.~\cite{edwards72} is determined by the matrix elements of the spin-current operator, \mbox{$j_{mn} = \langle n|j|m\rangle$},
\begin{equation}
R = \delta \varphi \sqrt{  \overline{ (\epsilon'_n(\varphi))^2  }}  /\Delta \epsilon= 
\delta \varphi  \; \sqrt{  \overline{ j^2_{nn}  }}  /\Delta \epsilon,
\quad \delta \varphi=\frac{\pi}{L}.
\label{rth}
\end{equation}
 Here, overline means averaging over the eigenstates 
 (in the relevant part of the spectrum) and $\Delta \epsilon$ is the average level spacing.
We note that integrable XXZ model 
with \mbox{$\Delta < 1$} exhibits  a ballistic transport. It is quantified 
by the spin stiffness (Drude weight)  \cite{zotos97}, $D=\overline{j^2_{nn} }/L > 0$ that is finite in the thermodynamics limit 
$L \to \infty$. Consequently, in ballistic models one obtains
$R= \pi \sqrt{D} /(\sqrt{L} \Delta \epsilon )$. 

In the case of the easy-axis regime, \mbox{$\Delta > 1$}, the ballistic component vanishes in GCE and in CE with $S^z_{tot}=0$. In both cases one finds $D=0$ for $L\to \infty$. However in the GCE, one may  still find a simple bound on the level sensitivity. To this end we employ the Mazur bound \cite{mazur69,zotos97},  $\overline{j^2_{nn} } \ge \langle j Q  \rangle ^2/\langle Q^2 \rangle$, where $Q$ is a conserved operator, \ch{
$\langle ... \rangle={\rm Tr}(...)/N_{st}$ and $N_{st}$ is the Hilbert-space dimension.}  We choose $Q$ as a product of two simplest local conserved operators, \mbox{$Q=Q_1 Q_3$}, where \mbox{$Q_1=\sum_l S^z_l$} is the total magnetization and  $Q_3$
is the energy current,
\begin{eqnarray}
 Q_{3} &=& \,\frac{iJ}{2} \sum_{l} \left( J S_{l+1}^{-} S_{l}^{z} S_{l-1}^{+} + \Delta S_{l+1}^{z} S_{l}^{+} S_{l-1}^{-} \right.\nonumber\\
 &&+ \left. \Delta S_{l+1}^{+} S_{l}^{-} S_{l-1}^{z} + \mathrm{H.c.} \right). 
\end{eqnarray}
Direct calculations in GCE give a bound
\begin{equation}
\overline{j^2_{nn} } \ge
\frac{\langle j Q_1 Q_3 \rangle ^2}{\langle Q^2_1 Q^2_3\rangle} = \frac{J^2\Delta^2 L}{2J^2(L-2)+4L\Delta^2},
\label{bound}
\end{equation}

We present in Fig.~\ref{fig1}(a) the GCE bound 
for different $\Delta \ge 1$ and compare it with the numerically evaluated
value $\overline{j^2_{nn}}$ obtained in GCE for systems with $ L \le 18$.
One can notice, that the lower bound even quantitatively approaches
the numerical value at $\Delta \gg 1$.
In the easy-axis regime, $\Delta > 1$, that is of interest here, one finds from Eq.~\eqref{bound} $\overline{j^2_{nn} }\ge 1/6$ 
and  $R\ge(0.4 \pi)/(L \Delta \epsilon )$.
In the GCE, the level sensitivity in ballistic and diffusive regimes of the integrable XXZ chains differ only by a factor \mbox{$1/\sqrt{L} \sim {\cal O}[\log(\Delta \epsilon)]$}.

The above bound is inapplicable in the CE with \mbox{$S^{z}_{tot}=0$} because $Q_1=0$. Fig.~\ref{fig1}(b) shows finite-size results for $\overline{j^2_{nn}}$. Here and further on, they are obtained using  exact diagonalization (ED).  For $L \le 22$, we carry out full diagonalization for sectors  with the wavevectors $0<q<\pi$  and  use half of the states from the middle of the spectra.  For the largest $L=24$ we employ the shift-invert ED and randomly sample over $N \sim 150$ eigenstates from the middle of the spectra, in each sector with $0<q<\pi$ (see Supplemental Material~\cite{supmat} for details).

\begin{figure}[h]
  \centering
  \includegraphics[width=\linewidth]{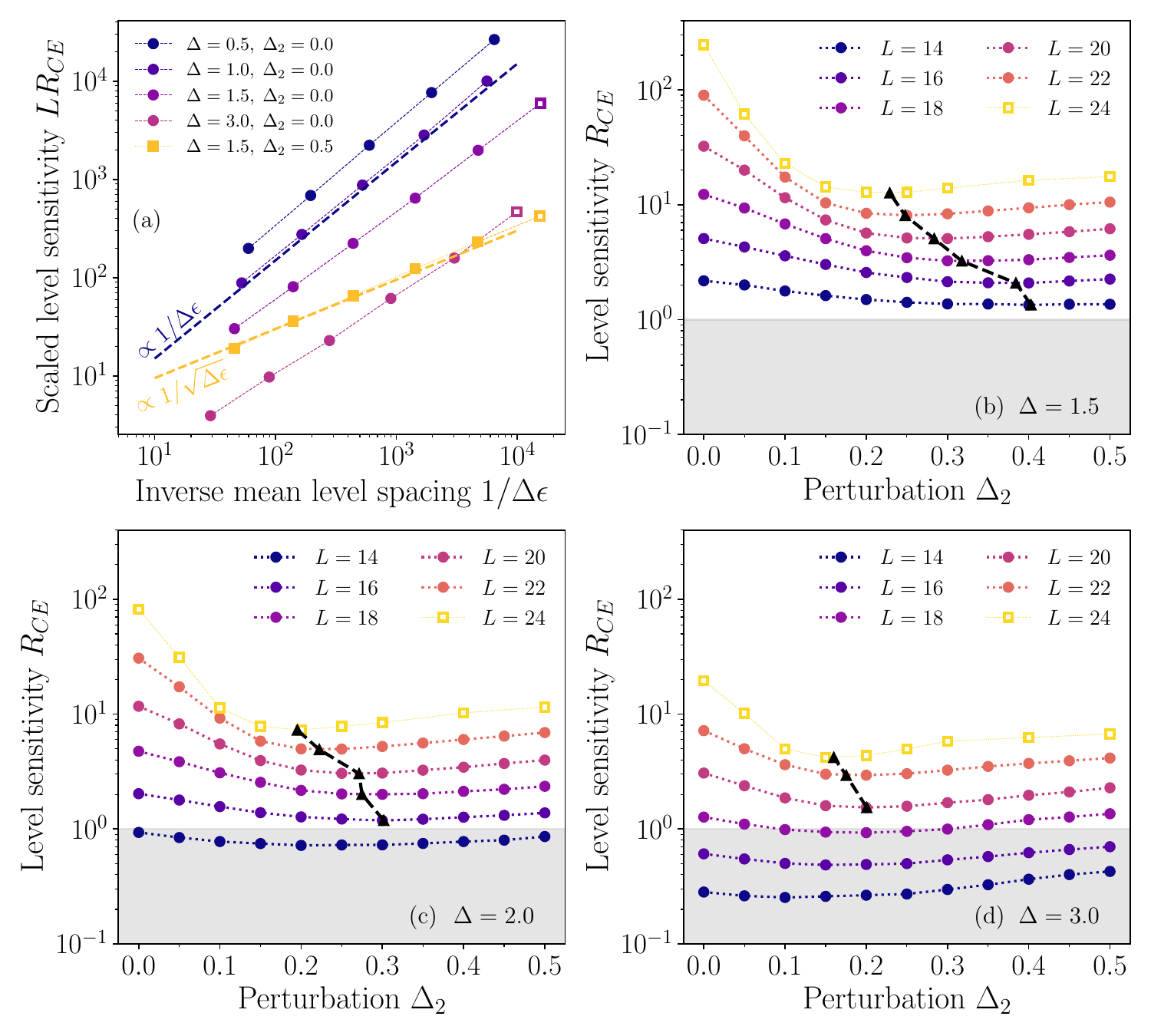}
  \caption{Results for CE with $S^z_{tot}=0$ from full ED (filled symbols) and shift-invert method (open symbols).
  (a) Scaled level sensitivity \(LR_{\text{CE}}\) vs \(1/\Delta \epsilon\), where $\Delta \epsilon$ is average level spacing in  sectors with fixed momenta. Dashed lines are guidelines showing $1/\Delta \epsilon$ and
  $1/\sqrt{\Delta \epsilon}$ dependence, respectively. (b)-(d) \(R_{\text{CE}}\) vs. IBP strength \(\Delta_2\) for \(\Delta=1.5, 2.0, 3.0\). \ch{Regions with $R<1$ are shaded, whereas black triangles correspond to minima of $R(\Delta_2)$, approximated from 
 a quadratic interpolation of three subsequent data points.}  
 }
  \label{fig2}
\end{figure}

One observes that  $\overline{j^2_{nn} }$  is well below the GCE bound and that the difference between GCE and CE results increases with $L$ and $\Delta$. In order to elucidate such $\Delta$-dependence, we have numerically studied also the 'folded' XXZ model \cite{zadnik21,zadnik21a,denardis22} that represents  $\Delta \to \infty$ limit of the XXZ chains,
\mbox{$H_{f} = (J/2) \sum_l P_l \left[ \mathrm{e}^{-i \varphi}
S^+_{l+1} S^-_l + \mathrm{H.c} \right]$} with the projection
operator $P_l = (S^z_{l+2}+S^z_{l-1})^2$ that introduces  conservation of the number of domain walls. In this model we find (so far numerically) strictly $\overline{j^2_{nn} }=0$ for any $L$. Consequently, one gets $R_{\text{CE}} =0$ while $R_{\text{GCE}} >0$. 
\ch{The latter follows from the strict bound in Eq.~\eqref{bound} and  
originates from the ballistic transport in all sectors with $S^z_{tot}\ne 0$.
Analytical support for vanishing $R_{\text{CE}}$ is discussed in the Supplemental Material. \cite{supmat}}
Following results in Fig.~\ref{fig1}(b) we present in Fig.~\ref{fig2}(a) the dependence of $R_{\text{CE}}$ on the average level spacing $\Delta \epsilon$. Results show that for finite 
$\Delta < \infty$, the anomalous dependence $R\propto 1/\Delta \epsilon$ indeed holds true, being thus close to ballistic systems rather than to generic diffusive ones.

The properties of $R$ in chaotic models are related to the d.c.~spin conductivity $\tilde{\sigma}$ via the RMT relations \cite{deutsch91,srednicki99,dalessio16}.
Since at high $T \gg J$ one gets $\tilde{\sigma} \propto T^{-1}$, we consider here
a rescaled $\sigma=T \tilde{\sigma}$, defined as the (Kubo) linear-response function
\begin{equation}
\sigma = \lim_{\omega \to 0}\frac{\pi}{L N_{st}} \sum_{n \ne m} 
| j_{mn} |^2 \delta(\omega-\epsilon_m + \epsilon_n)
\simeq \frac{ \pi\overline{ | j_{mn}|^2}}{L  \Delta \epsilon} ,  \label{dom}
\end{equation}
where $\overline{ | j_{mn}|^2}$ are averaged matrix elements for small $|\epsilon_m-\epsilon_n|$.
Upon introducing a  flux $\varphi \ne 0$, the degeneracy of energy levels persists only in momentum sectors $q=0$ and $q=\pi$. All other sectors  obey the RMT relations~\cite{dalessio16}, 
 which in our system  follow the Gaussian orthogonal ensemble (GOE) universality, requiring $  Y = \overline{ | j_{mn}|^2}/ \overline { j_{nn}^2 } = 1/2$. Therefore, RMT establishes a link between the level sensitivity, Eq.~\eqref{rth} and the spin conductivity, Eq.~\eqref{dom},
 \begin{equation}
 R=\sqrt{\frac{2 \pi \sigma}{L \Delta \epsilon}}.
 \label{rth1}
 \end{equation}

\ch{Eq. (\ref{rth1}) means that the level sensitivity in chaotic systems scales with the system size as $R\propto 1/\sqrt{L \Delta \epsilon}$.  It is a general (i.e. model independent) property that follows solely from RMT and the finiteness of $\sigma$. In the case of integrable systems, $R$ should follow a simple bound $R \ge c/(L^{m+1} \Delta \epsilon)$ where $c$ is a constant of order one.  Independently of the studied model, a sufficient condition for the latter to hold is that the current operator has a non-vanishing projection on products of a finite number of LIOMs, $\langle jQ_1Q_2\ldots Q_m \rangle\ne 0$, see the discussion in the Supplemental Material \cite{supmat}. We stress that the latter projection implies ballistic transport only for $m=1$. We also recall that the level spacing, $\Delta \epsilon$, decays exponentially with $L$. Consequently, introducing of IBP  should in general be accompanied by a substantial drop of the level sensitivity from \mbox{$R\propto 1/\Delta \epsilon$} in the integrable chain (either ballistic or diffusive) to $R\propto 1/\sqrt{\Delta \epsilon}$ in the chaotic system. All other terms which enter $R$  can be bounded by a power-law function $\sim 1/L^{m+1}$ so that they appear only as logarithmic corrections, $\sim \log(\Delta \epsilon)$.}

\begin{figure}[h]
  \centering
  \includegraphics[width=\linewidth]{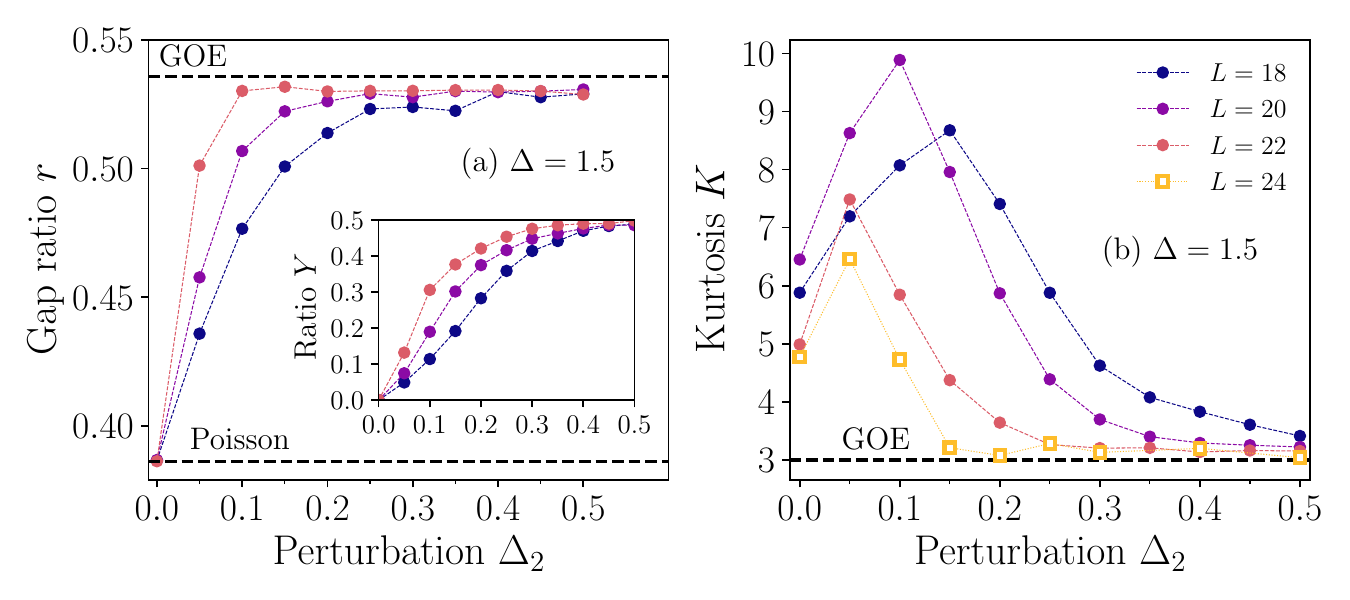}
  \caption{RMT indicators for the perturbed model in canonical sector. a) Gap ratio \(r\), inset: ratio $Y$ of off-diagonal and diagonal  matrix elements. b) Kurtosis $K$ (defined in text).}
  \label{fig3}
\end{figure}

\ch{All  result henceforth apply to CE.
Figure~\ref{fig2}(a)  confirms that $R_{\text{CE}}$ in the integrable XXZ model shows the dependence $R_{\text{CE}} \propto 1/\Delta \epsilon$ for all finite $\Delta$. Comparing this result with
$R_{\text{CE}}=0$ in the 'folded' model, one finds that the limits $L\to \infty$ and $\Delta \to \infty$ do not commute. The limit $\Delta \to \infty$ is taken first in the 'folded' model.   
Fig.~\ref{fig2}(a)  confirms also} that substantial IBP ($g=\Delta_2$) causes qualitative changes of $R_{\text{CE}}$, as discussed in the preceding paragraphs. More systematic dependence 
of  $R_{\text{CE}}$ on the IBP strength $\Delta_2$ is presented in Figs.~\ref{fig2}(b)-~\ref{fig2}(d) for different 
$\Delta > 1$.
One observes that $R_{\text{CE}}$ develops a clear minimum at finite $\Delta_2$ that drifts towards weaker IBP  when $L$ increases. 
It should be noted, that we are interested in the regime $R_{\text{CE}} \gtrsim 1$. For $R_{\text{CE}} \lesssim 1 $ (shaded areas in Fig.~\ref{fig2}) the system is too small, i.e.~the level spacing is too large, so that changing of the flux causes neither level crossings nor avoided crossings (within the same $S^z_{tot}$ and momentum sector).

\begin{figure}[h]
  \centering
  \includegraphics[width=\linewidth]{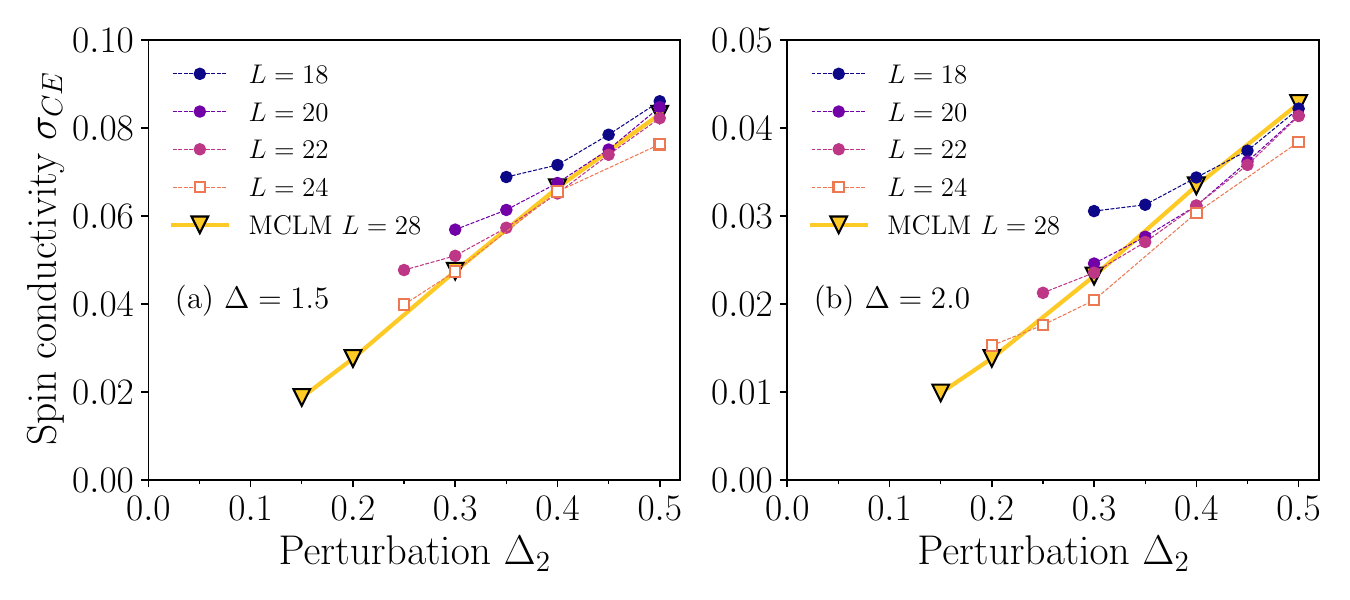}
  \caption{Spin conductivity in canonical sector,  \(\sigma_{\text{CE}}\),  for: (a) \(\Delta=1.5\) and (b) \(\Delta=2.0\). Results obtained from diagonal matrix elements via full ED (\(L=18 - 22\)) and shift-invert (\(L=24\)) are presented in the regime $\Delta_2 > \Delta_2^*(L)$, i.e., above the crossover to the GOE universality. Presented are also MCLM results for
  $L=28$.} 
  \label{fig4}
\end{figure}

In order to explain the origin of this minimum,
we  numerically study the standard 
indicators of RMT. 
The main panel in Fig.~\ref{fig3}(a) shows the gap ratio $r =  \overline r_n $ with \mbox{$r_n =\mathrm{min} [\Delta_n,\Delta_{n+1}]/\mathrm{max}[\Delta_n,\Delta_{n+1}]$}, where \mbox{$\Delta_n=\epsilon_{n+1}-\epsilon_n$}. The inset shows the ratio of the off-diagonal and the diagonal matrix elements, \mbox{$Y=\overline{ | j_{mn}|^2}/\overline { |j_{nn}|^2 }$}. Both results confirm that sufficiently strong IBP introduces the GOE universality \cite{oganesyan07,atas13} with $Y=0.5$ and $r\simeq 0.53$.
Fig.~\ref{fig3}(b) shows  the kurtosis $K= \overline{j_{nn}^4} / \left(\overline{j_{nn}^2}\right)^2$. At large enough $\Delta_2$, $K$ agrees with the kurtosis of the Gaussian distribution with $K=3$. Comparing results in Figs.~\ref{fig2} and~\ref{fig3} one finds that minimum in $R_{\text{CE}}$  corresponds to the  crossover to the normal RMT behavior. In other words, the minimum reveals the threshold strength of IBP, $g^*(L)$, when the  scattering becomes strong enough to introduce the RMT universality.

In systems which satisfy  RMT, the  spin conductivity, $\sigma$, can be  obtained also from the diagonal matrix elements,
$\overline { |j_{nn}|^2 }$, via Eqs.~\eqref{rth} and~\eqref{rth1}.  Fig.~\ref{fig4} 
shows such results for the spin conductivity in CE,  $\sigma_{\text{CE}}$.
As before we use $g = \Delta_2$  as IBP. In Fig.~\ref{fig4}  we present only numerical results which are beyond the crossover to the GOE universality, i.e. obtained for $g > g^*(L)$. The minima in $R_{\text{CE}}$ in Figs.~\ref{fig2}(b)-~\ref{fig2}(d) are here visible as the starting points of curves obtained for various L. In Fig.~\ref{fig4} we 
show also results obtained from the microcanonical Lanczos method (MCLM) \cite{long03,prelovsek13}. This method allows to evaluate full dynamical $\sigma_{\text{CE}}(\omega)$, and consequently also $\sigma_{\text{CE}}$ as in Eq.~(\ref{dom}), for larger systems, but with smaller frequency resolution $\delta \omega$.
Here, we present MCLM results for $L=28$ with $\delta \omega \lesssim 10^{-3} $ \cite{prelovsek22}, which we estimate to be  meaningful  at least down to $g=\Delta_2 \sim 0.15$. 

 We note that results for $\sigma_{\text{CE}}$ obtained from various methods are  consistent provided that $g > g^*(L)$. One also observes that $\sigma_{\text{CE}}$  strongly depends on $\Delta$ resembling the analogous dependence of $R_{\text{CE}}$ on $\Delta$ shown   Fig.~\ref{fig2}(a). Apparently for $g>g^*(L)$,  the spin conductivity  follows the relation $\sigma_{\text{CE}} \propto |g| $. Here we can make a link with the diffusion constant, ${\cal D}$. In generic dissipative system it is related to $\sigma$
via the Einstein relation, that at large $T$ reads  ${\cal D} = 4 \sigma$. The diffusion has been studied extensively in the integrable case and evaluated mostly in the GCE, both analytically within the GHD \cite{gopalakrishnan19}  and numerically via various methods \cite{prelovsek04,znidaric11,steinigeweg15,karrasch14,prelovsek22} leading
to values ${\cal D}_{\text{GCE}} > 0.4$ in the regime $\Delta > 1 $ . Evidently,
$\sigma_{\text{CE}}$ shown in Fig.~\ref{fig4}  for perturbation $g>g^*(L)$ is well below the latter.
This discrepancy suggests a
discontinuous change of ${\cal D}$ upon the introduction of IBP in agreement with Refs. \cite{denardis22,prelovsek22}.

{\it Conclusions and Discussion} We have studied sensitivity of the energy levels to the flux (or equivalently twisted boundary conditions), $R$, and related transport quantities, within the XXZ chain in the easy-axis regime. Although the transport in the integrable model is diffusive, obtained $R$ is very close to one in integrable ballistic systems and, up to logarithmic corrections, exhibits the same dependence on the level spacing, $R\propto 1/\Delta \epsilon$. It strongly contrasts with chaotic diffusive systems where $R\propto 1/\sqrt{\Delta \epsilon}$. In the case of GCE , we have found for integrable chain a strict bound $R_{\text{GCE}} \ge 0.4 \pi/(L \Delta\epsilon)$ that holds for arbitrary anisotropy $\Delta \ge 1$ and all system sizes, $L$. In the CE and for finite anisotropy $\Delta < \infty$, the level sensitivity still varies as $R_{\text{CE}}\propto 1/\Delta \epsilon$, however $R_{\text{CE}}$ is much smaller than $R_{\text{GCE}}$. The difference between  $R_{\text{GCE}}$ and $R_{\text{CE}}$ grows with the systems sizes and anisotropy. In the limiting case $\Delta \to \infty$ we have found that $R_{\text{CE}}$ strictly vanishes for all system sizes whereas  $R_{\text{GCE}}$  is bounded from below. \zm{In the Supplemental Material \cite{supmat}, we discuss results for nonzero magnetization sectors.}

The qualitative difference between the level sensitivity in integrable and chaotic systems implies that a  perturbation of strength $g$ must cause a substantial reduction of  $R$.  In finite systems,  $R_{\text{CE}}$ is minimal at the crossover between integrable and RMT regimes.  The minimum at $g=g^*(L)$ shifts
towards weaker perturbations with increasing L. 

In the chaotic regime,  RMT establishes the Thouless-like relation between $R$ and the d.c.~spin conductivity, $\sigma$. Using this relation we have numerically studied  $\sigma$ in CE and found for $g>g^*(L)$ that  $\sigma_{CE} \propto |g|$. For weak perturbations,  $\sigma_{CE}$ becomes much smaller than $\sigma_{GCE}$ estimated from the diffusion constant for integrable systems.  
Assuming that the Einstein relation and the equivalence
between CE and GCE hold true for all model parameters, our results support a discontinuous variation of  $\sigma_{GCE}=\sigma_{CE}$ with IBP. It should be acknowledged that the limit of a weak perturbation is highly nontrivial since the difference between CE and GCE results goes as, $1/(g^2L)$ \cite{prelovsek22}. Consequently, the convergence  of results in both ensembles is hard to establish numerically.

\nocite{zenodo}
\vskip 1truecm  
\noindent {\it Acknowledgments.} We acknowledges fruitful discussions with L. Zadnik and R. Steinigeweg.
M.M. acknowledges support by the National Science Centre (NCN), Poland via project 2020/37/B/ST3/00020. J.P. acknowledges support by the National Science Centre (NCN), Poland via project 2023/49/N/ST3/01033. P. P. acknowledges
support of the Slovenian Research Agency via the program P1-0044.

\bibliography{manupert}
\newpage
\phantom{a}
\newpage

\setcounter{figure}{0}
\setcounter{equation}{0}
\setcounter{page}{1}

\renewcommand{\thetable}{S\arabic{table}}
\renewcommand{\thefigure}{S\arabic{figure}}
\renewcommand{\theequation}{S\arabic{equation}}
\renewcommand{\thepage}{S\arabic{page}}

\renewcommand{\thesection}{S\arabic{section}}

\onecolumngrid

\begin{center}
{\large \bf Supplemental Material:\\
Thouless approach in transport in integrable and perturbed easy-axis Heisenberg chains}\\
\vspace{0.3cm}
J. Pawlowski $^{1}$, M. Mierzejewski$^{1}$, and P. Prelov\v{s}ek$^{2}$ \\
$^1$ {\it Institute of Theoretical Physics, Faculty of Fundamental Problems of Technology, Wroc\l{a}w University of Science and Technology, 
50-370 Wroc\l{a}w, Poland}\\
$^2${\it J. Stefan Institute, SI-1000 Ljubljana, Slovenia} \\
\end{center}

In the Supplemental Material we present  additional results with next-nearest-neighbor exchange as the perturbation, \ch{ details of shift-invert calculations, and extra considerations of the 'folded' XXZ model, which can serve as the toy model for the behavior of the easy-axis $\Delta > 1$ regime. We also show an analytical argument bounding the diagonal matrix elements of the spin current.} \zm{Finally, we discuss the level sensitivity for nonzero magnetization sectors.}
 
\vspace{0.6cm}
\twocolumngrid

\label{pagesupp} 
\subsection{Next-nearest-neighbor exchange as the perturbation: results}

In addition to the next-nearest-neighbor (nnn) interaction $g=\Delta_2$ taken as the integrability-breaking perturbation (IBP) we present 
analogous results also for the 
the nnn exchange where $g=J_2$ and 
\begin{equation}
H^\prime = \sum_l  [\mathrm{e}^{-2 i \varphi} S^+_{l+1} S^z_l S^-_{l-1} +  \mathrm{H.c.}].
\end{equation} 
We note that such term  maps to a nnn hopping in the fermionic  chain. 

\begin{figure}[h]
  \centering
  \includegraphics[width=\linewidth]{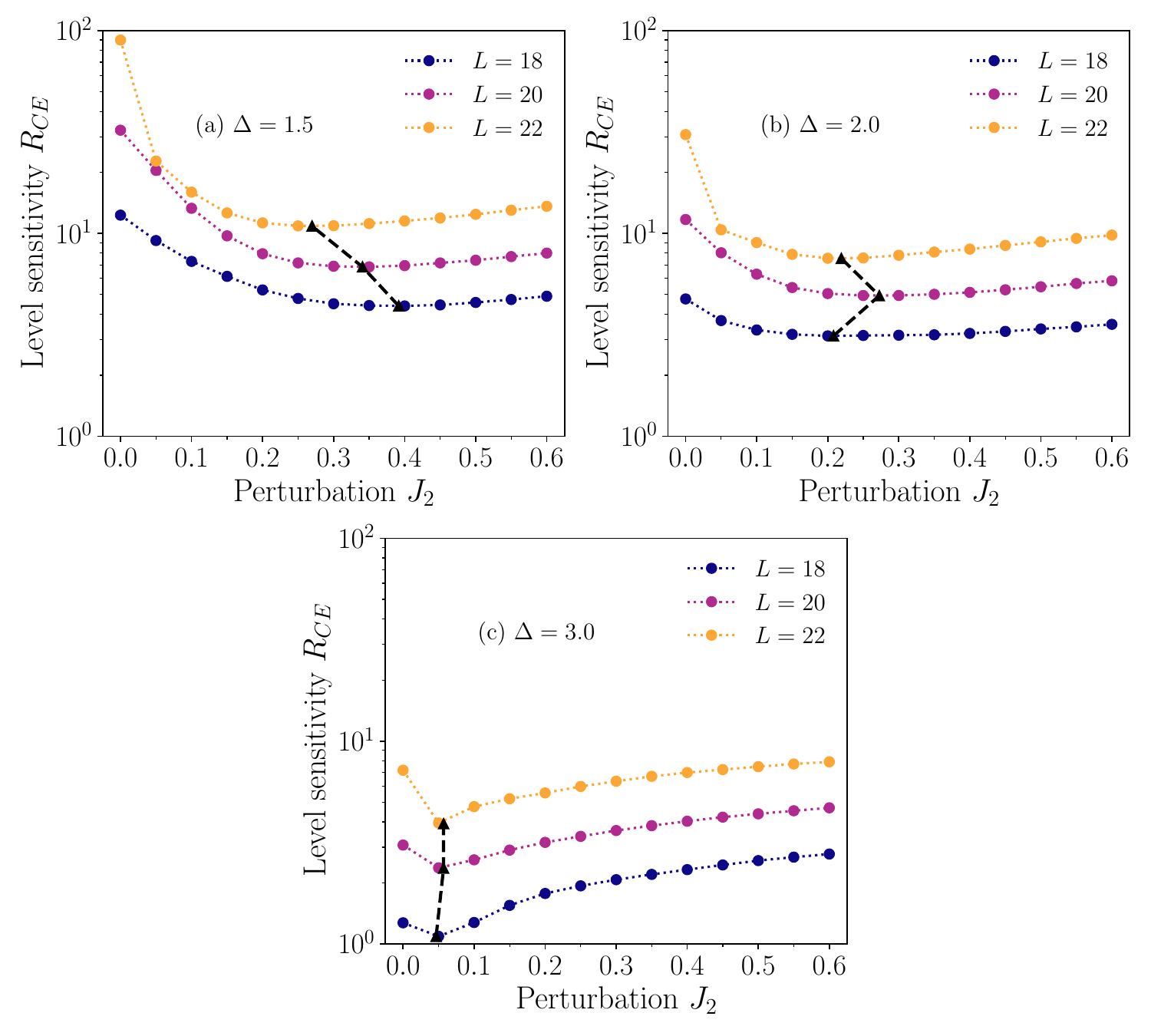}
  \caption{Level sensitivity $R_{CE}$ vs. perturbation strength $J_2$ within sector $S^z_{tot}=0$, as calculated via full ED in systems with $L = 18 -22$ sites, for: (a) $\Delta = 1.5$, 
  (b) $\Delta = 2.0$, and (c) $\Delta= 3.0$. \ch{Black triangles correspond to minima of $R(\Delta_2)$,  approximated from 
 a quadratic interpolation of three subsequent data points.}}
  \label{supfig1}
\end{figure}

We first present in Fig.~\ref{supfig1} results for the level sensitivity $R$ varying the IBP strength $J_2$, as calculated via full ED in systems with $ L = 18 - 22$. Again,
substantial IBP ($J_2 \gg 0$) causes qualitative changes of $R$, inducing a minimum of $R(J_2)$ indicating the crossover into the regime $J_2 > J_2^*(L)$ 
with RMT universality. As in the main text, also here we restrict to the regime $R>1$ where the interpretation of results in terms of normal transport is meaningful. While the behavior is quite similar to the case of $g = \Delta_2$,
 presented in the main text, it should be noticed that the dependence of $R$ on $\Delta$ in Fig.~\ref{supfig1} is less pronounced  than for  $\Delta_2$.  Larger values of $R$ in the present case allow us also to reach $R>1$ also for $\Delta =3$. The qualitative difference emerges at $\Delta \to \infty$ where
in the folded model with $\Delta_2 \ne 0$ one  obtains $R = 0$, while this is not the case for
$g = J_2$  which introduced $R>0$ even in the folded model.

\begin{figure}[h]
  \centering
  \includegraphics[width=\linewidth]{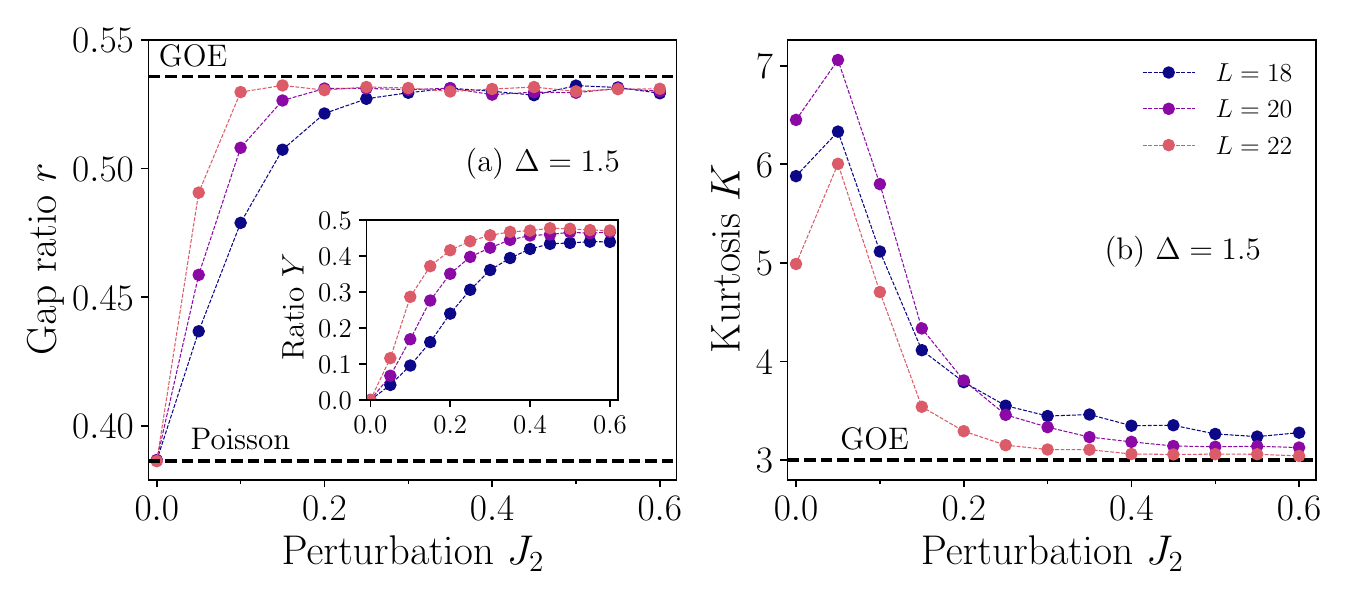}
  \caption{RMT indicators for the model with \(J_2\) perturbation. a) Gap ratio \(r\), inset: ratio $Y$ of off-diagonal matrix elements versus the diagonal ones. b) Kurtosis \(K\) (defined in the text).}
  \label{supfig2}
\end{figure}

We present in  Fig.~\ref{supfig2} results for standard
indicators of RMT: in  Fig.~\ref{supfig2}(a) the gap ratio $r =  \overline r_n $ with \mbox{$r_n =\mathrm{min} [\Delta_n,\Delta_{n+1}]/\mathrm{max}[\Delta_n,\Delta_{n+1}]$}, in the inset of Fig.~\ref{supfig2}(a) the ratio \mbox{$Y=\overline{ | j_{mn}|^2}/\overline { |j_{nn}|^2 }$}, and finally in  Fig.~\ref{supfig2}(b) the kurtosis $K= \overline{j_{nn}^4} / \left(\overline{j_{nn}^2}\right)^2$. All results again confirm that sufficiently strong IBP
$g > g^*(L)$ introduces the GOE universality with $Y=0.5$, $r\simeq 0.53$, and $K \simeq 3$.

\begin{figure}[h]
  \centering
  \includegraphics[width=\linewidth]{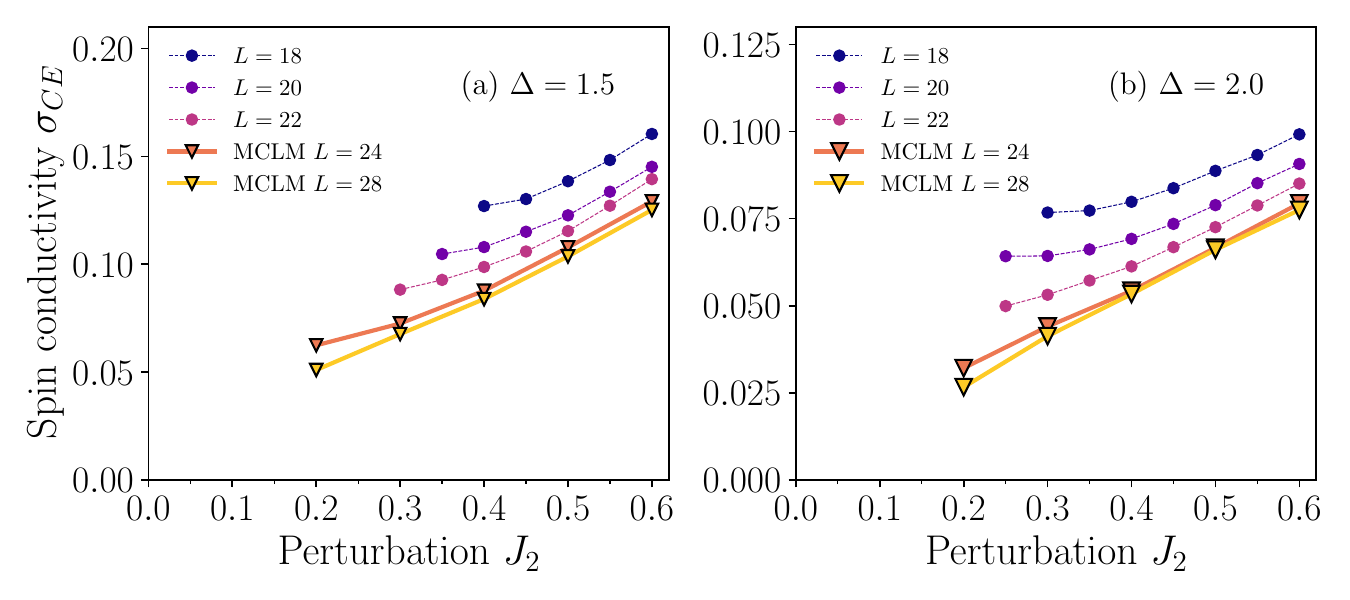}
  \caption{Spin conductivity \(\sigma_{CE}\) for (a) \(\Delta=1.5\) and (b) \(\Delta=2.0\). Points denote results obtained from diagonal matrix elements via full ED (\(L=18 - 22\)), presented in the regime $J_2 > J_2^*(L)$, i.e., above the crossover to the GOE universality. Presented are also MCLM results for $L=24$ and $L=28$.} 
  \label{supfig3}
\end{figure}

Finally, we show results for $\sigma_{CE}$ beyond the crossover to the GOE universality $g > g^*(L)$ for $g = J_2$, here obtained via ED with up to $L=22$. Again, results for $\sigma_{CE}$ become quantitatively consistent (weakly $L$-dependent) provided that $g > g^*(L)$.  Extrapolating the crossover $g^*(L)$ to larger $L$, we can compare then results to those obtained via MCLM  for  larger $L =24$ and $L=28$, which then apparently leads  to RMT-meaningful results down to $g=J_2 \sim 0.2$ as presented in Fig.~\ref{supfig3}. The conclusion are  analogous to one presented for $g = \Delta_2$. The difference again is that  the dependence 
of $\sigma_{CE}$ on $\Delta $ for $g = J_2$ is weaker than for $g = \Delta_2$. Namely, $\sigma_{CE}$ at $J_2>0$
remains substantial  also in the limit $\Delta  \to \infty$ , i.e., within the folded model. Otherwise, for $J_2 \ll 0.5$ the values of $\sigma_{CE}$ are well below the integrable GCE value $\sigma_{GCE} > 0.1$  indicating  on the discontinuous jump at $g \to 0$. Moreover,  
the variation of $\sigma_{CE}$ with $g$ in Fig.~\ref{supfig3} reveals $\sigma_{CE} \propto |g| $ for $g>g^*(L)$. However, as in the main text we  stress that the regime of weak IBP should be taken with care and requires further attention.

\subsection{Numerical methods}

We evaluate the diagonal matrix elements of the spin current for \(L=24\) using a custom implementation of the shift-invert method~\cite{pietracaprina18}, employing Intel oneMKL for matrix-vector multiplication and oneMKL PARDISO linear solver to compute the action of \((H-E)^{-1}\) on vectors. The essence of this approach is to apply the standard Lanczos algorithm to the resolvent \((H-E)^{-1}\), where \(E\) is a target energy, in our case selected randomly from the middle half of the spectrum.
\begin{figure}[h]
  \centering
  \includegraphics[width=\linewidth]{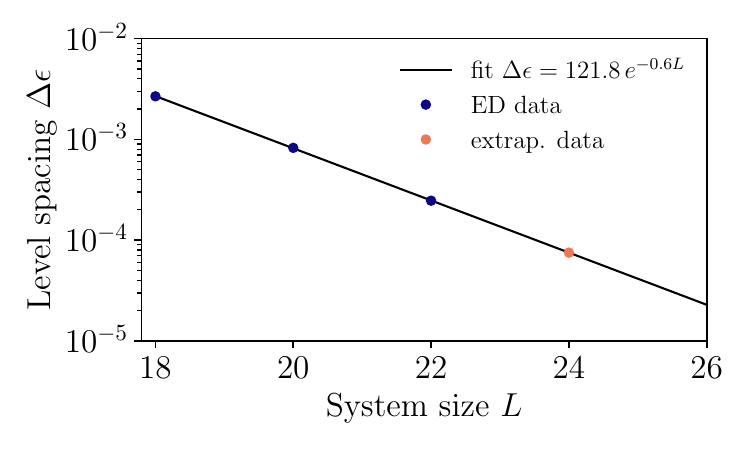}
  \caption{Mean level spacing \(\Delta \epsilon\) vs. system size \(L\) for \(\Delta=2.0\) and \(\Delta_2 = 0.1\). Dark blue points correspond to ED data for \(L=18,20,22\) and orange point to the extrapolated value for \(L=24\), used in subsequent shift-invert calculations.} 
  \label{supfig4}
\end{figure}
Using the resolvent of the Hamiltonian turns a dense part of the spectrum into a sparse one, facilitating the convergence of Lanczos procedure. In this way, we can sample the relevant eigenstates and approximate moments of \(j_{nn}\) without performing a full diagonalization. We find \(M=30\) Lanczos steps and \(100-150\) states per momentum sector being enough for the results to converge. However, there is some caution required, as the action of the resolvent on a vector results in a system of linear equations \((H-E)|w\rangle = |v \rangle \), that is severely ill-conditioned, with condition number \(\kappa \propto \exp{L}\). Hence, iterative solvers usually fail to reach satisfying precision, and direct solvers have to be used instead, limiting accessible systems sizes.

To compute the level sensitivity \(R\)  and spin conductivity \(\sigma_{\text{CE}}\) (Eq.~\eqref{rth} and Eq.~\eqref{dom} in the main text), we also require the mean level spacing \(\Delta \epsilon\) in the central part of the spectrum. Taking into account the exponential dependence of \(\Delta \epsilon\) on the size of the system, it is straightforward to extrapolate this value from the ED data for \(L=18,20,22\). An example of such an extrapolation for the case of \(\Delta = 2.0\) and \(\Delta_2 = 0.1\) is shown in Fig.~\ref{supfig4}.

\subsection{Folded XXZ model and its properties as the toy model for the easy-axis regime}

\ch{The 'folded' XXZ model can be regarded as the easy-axis limit, $\Delta \to \infty$, of the
XXZ chain. In order to calculate the matrix elements of the spin current operator, $j_{nn}$, we study this model in the presence of a nonzero flux $\varphi$,
\begin{equation}
\tilde H = \frac{J}{2} \sum_l  P_{l} [~\mathrm{e}^{-i \varphi} S^+_{l+1} S^-_l + \mathrm{H.c.} ~],
\label{fold}
\end{equation}
Here, the projector, $P_{l} = [S^z_{l+2} + S^z_{l-1}]^2$, ensures that the spin exchange does not modify the total number of domain walls (DW).  The GCE the lower bound, presented in the main text in Eq. (\ref{bound}) is valid also for the 'folded' model,
i.e. $\overline{ j_{nn}^2}= 1/4$ and appears to be tight in this limit (see Fig.~\ref{fig1}). On the other hand,
the CE results appear to be strictly zero, $\overline{ j_{nn}^2} = 0$, (at least numerically) for any finite $L$, 
for which we give further analytical arguments and explanation. 
}

\ch{The aim is to present arguments that there are no diagonal current elements $j_{nn}$ in any eigenstate 
at $S^z_{tot}=0$, sticking to systems with even $L$. It is convenient to start from the reference antiferromagnetic (AFM) state and subsequently add additional domain walls $N_W = 0, 2, 4 \ldots$.  While in the $N_W =0$ sector there is 
no  dynamics within the model, Eq.~\eqref{fold}, the first nontrivial subspace is $N_W = 2$.}

\begin{figure}[htbp]
    \centering
    \includegraphics[width=0.8\linewidth]{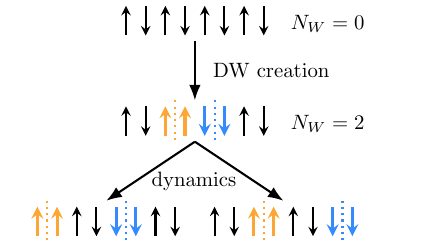}
 \caption{\ch{Sketch showing dynamics of the 'folded' model with $N_W=2$ domain walls (marked with dotted lines) in the antiferromagnetic background.} }
 \label{sm_nw2}
\end{figure}

\ch{The pair of DW can move and behave as charged solitons (particles), see Fig.~\ref{sm_nw2}. For $N_W=2$ they can be described by the toy model,
\begin{equation}
H_{\text{t}} = t \sum_l  [ ~\mathrm{e}^{i \varphi} d^\dagger_{l+2} d_l + \mathrm{e}^{-i \varphi} c^\dagger_{l+2} c_l
 + \mathrm{H.c.} ~], \label{heff}
 \end{equation}
where $c_l,d_l$ represent, respectively, positive and negative solitons (DW) on a bond between sites  $l$ and $l+1$, and \mbox{$t =J/2$}.
DW move on one sublattice
(even or odd), but are impenetrable $ d^\dagger_{l} d_l c^\dagger_l c_l = 0$ and cannot cross each other.
The two-particle problem can be solved exactly, but here the key feature is that the energies are independent of $\varphi$,
hence $j_{nn} = {\rm d} \epsilon_n(\varphi)/{\rm d} \varphi =0$. This can be confirmed in different ways. 
The eigenproblem for the translationally-invariant toy model with periodic boundary conditions can be written down in the general form, 
\begin{equation}
| \Psi \rangle =  \sum_m  f_m |\tilde \psi_{m} \rangle,
\quad |\tilde \psi_m \rangle = \sum_{j}  \mathrm{e}^{i 2 K j } T_{ 2j} c^\dagger_{2 m+1} d^\dagger_1 | 0  \rangle,
\end{equation} 
where $K$ is a wave-vector and $T_{2j}$ is the translation operator that shifts 
positions of both particles by $2j$ sites. The latter translation takes into account that in the toy model,
Eq.~\eqref{heff}, DW hops by two sites, that is, DW stays on the same sublattice.
 In the new basis the Hamiltonian acts as
\begin{eqnarray}
&&H_{\text{t}} | \tilde \psi_1\rangle =   t ~\mathrm{e}^{i \varphi} (1 +  \mathrm{e}^{i 2 K} ) |\tilde \psi_2\rangle~, \nonumber \\
&&H_{\text{t}} | \tilde \psi_l\rangle =   t ~\mathrm{e}^{-i \varphi} (1 +  \mathrm{e}^{-i 2 K} ) |\tilde \psi_{l-1}\rangle  + t ~\mathrm{e}^{i \varphi} 
(1 +  \mathrm{e}^{i 2 K} ) |\tilde \psi_{l+1}\rangle~, \nonumber \\
&&H_{\text{t}} | \tilde \psi_{L/2-1}\rangle =   t ~\mathrm{e}^{-i \varphi} (1 +  \mathrm{e}^{-i 2 K} ) |\tilde \psi_{L/2-2}\rangle~.
\end{eqnarray}
The resulting Hamiltonian matrix is tri-diagonal, where all non-diagonal elements are equal $b_l = t ~\mathrm{e}^{i \varphi} (1 +  \mathrm{e}^{i 2 K} )$.
Most importantly, the eigenenergies, $\epsilon_n$, are determined by the products $B_l  = b_l b^*_l $ that are real and do not involve $\varphi$.
Hence, $\epsilon_n$ are independent of $\varphi$ and consequently $j_{nn} =0$. }

\begin{figure}[htbp]
    \centering
    \includegraphics[width=0.8\linewidth]{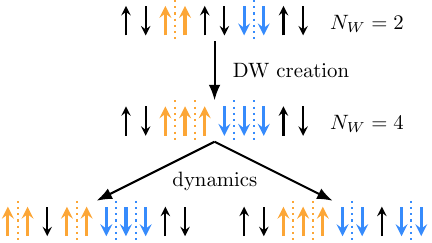}
     \caption{\ch{The same as in Fig. \ref{sm_nw2} but with $N_W=4$ domain walls.} } 
  \label{sm_nw4}    
\end{figure}

\newpage
\ch{In the same way, one can introduce additional DW, keeping  $S^z_{tot}=0$. In particular, the case with 
 $N_W =4$ domain walls is sketched  in Fig. \ref{sm_nw4}. }

\ch{In the above case we introduced solitons (DW) both on even and odd sites, and they can again move by hopping
by two sites. They are clearly also other possibilities to generate
$N_W > 2$.  It is important that positive and negative solitons 
are always created in pairs and that  they are effectively impenetrable/noncrossing
being the origin of $j_{nn}=0$, confirmed numerically for all configurations.}

\ch{It is relevant to point out the analogy of physics of 'folded' model and Eq.~(\ref{heff})  to the $t$ model \cite{prelovsek04}, i.e., the
Hubbard model at $U \to \infty$ with magnetic flux (acting oppositely on up/down spins),
\begin{equation}
H_{\text{im}} = t \sum_l  [ ~\mathrm{e}^{-i \varphi} c^\dagger_{ l+1\uparrow } c_{ l\uparrow} + \mathrm{e}^{i \varphi} 
c^\dagger_{l+1\downarrow} c_{ l\downarrow}  + \mathrm{H.c.} ~], \label{him}
 \end{equation}
 with imposed  hard-core $U \to \infty$ repulsion $n_{\downarrow l}  n_{\uparrow l}=0 $. This model has the same properties as the 'folded' model with respect to the spin current $j_{nn}= {\rm d} \epsilon_n(\varphi)/{\rm d} \varphi = 0$, provided that $N_\uparrow = N_\downarrow$ \cite{prelovsek04},  
whereby in Eq.~(\ref{him}) the 
structure of (only nearest-neighbor) hopping is simpler.}

\ch{Finally, the analysis of the `folded` model has the relevant consequence even for perturbed systems at
$\Delta \to \infty$. Namely, the introduction of the next nearest neighbor interaction, $\Delta_2 > 0$,  influences neither the
impenetrability of DW nor the independence of energies on the flux, $\epsilon_n (\varphi)= \epsilon(0)$. Consequently one gets $j_{nn}=0$
also for the perturbed `folded` model. It 
explains why we also find numerically that $\sigma_{CE}$ decreases with $\Delta$ even at finite
$\Delta_2$. On the other hand, $g=J_2$ perturbation directly  allows for crossing of DW,
hence even $\Delta \to \infty$ limit is non-trivial and the perturbed `folded` model can serve as
the nontrivial starting point.}

\subsection{Bound on the diagonal matrix elements of the spin current} 
\ch{
In the main text, we introduced a product of two LIOMs, \(Q=Q_1 Q_3\), and used it to prove
an explicit bound on the diagonal matrix elements $\overline{j^2_{nn}}$ in the easy-axis XXZ chain, see Eq. (\ref{bound}). In what follows, we argue that similar reasoning can be generalized for other translationally invariant, integrable models. Namely, we demonstrate that if the spin current has nonvanishing projection on any product of $m$ LIOMs,
\begin{equation}
\langle j Q_1Q_2\ldots Q_m \rangle \ne 0,
\label{assums}
\end{equation}
then the Mazur-bound can be expressed in terms of a power-law dependence on $L$
\begin{equation}
\overline{j^2_{nn} } \ge
\frac{\langle j Q_1 Q_2\ldots Q_m \rangle ^2}{\langle Q^2_1 Q^2_2\ldots Q^2_m\rangle} \ge \frac{\rm const}{L^{2m}}.
\label{bounds}
\end{equation}
 Here,  $Q_i$, $i=1,\ldots,m$, denote {\em arbitrary} LIOMs.
In order to demonstrate the inequality on  the right-hand-side of 
Eq.~(\ref{bounds}), we  note that the normalization of $Q_i$
is not important. One may multiply any LIOM by a constant, $Q_i \to c_i Q_i$, and the constant $c_i$ cancels out in Eq.~(\ref{bounds}). Therefore, we assume that $Q_i$ are translationally invariant and {\em extensive} operators, $Q_i=\sum_{l=1}^L (Q_i)_l$.  Here, $(Q_i)_l$ is the local density of $Q_i$ supported on a few sites in the vicinity of site $l$.  Then, the term in the numerator in Eq. (\ref{bounds}) reads
\begin{equation}
\langle j Q_1... Q_m \rangle=
\sum_{l_0=1}^L \sum_{l_1=1}^L \ldots \sum_{l_m=1}^L
\langle (j)_{l_0} (Q_1)_{l_1}\ldots (Q_m)_{l_m} \rangle.
\label{boundss}
\end{equation}
In the case of extensive operators, the corresponding densities $(Q_i)_l$ and $(j)_l$ are independent of the system size. Therefore, all summands in Eq.~\eqref{boundss} are also 
\mbox{$L$-independent} and they are either  zero or are of the order one ($\sim L^0$). Due to the assumption posed by Eq.~\eqref{assums},
some terms must be nonzero. If a certain term is nonzero then, due to translational invariance, there are $L$ identical summands in Eq.~\eqref{boundss}. Consequently, the left-hand side of Eq.~\eqref{boundss} should be (at least) of the order $L$, unless there is some peculiar cancellation of positive and negative summands. However for the  discussion of the level sensitivity, one may safely assume that $\langle j Q_1\ldots Q_m \rangle$ is of order $L$. The denominator in Eq.~\eqref{bounds} represents  squared norm of the operator $Q_1\ldots Q_m$ so it is positive and nonzero. One may write down $\langle Q^2_1... Q^2_m \rangle$ in the form analogous to Eq.~\eqref{boundss}. The resulting expression contains summations of terms which are either zero or are of order one and summations are carried out over $2m$ lattice indexes. We use the  simplest bound  $\langle Q^2_1\ldots Q^2_m \rangle \le c L^{2m} $, where $c$ is a constant of the order one. Consequently, one obtains the bound from Eq.~\eqref{bounds}, $\overline{j^2_{nn} } \ge {\rm const}/L^{2m}$.  This combined together with the definition of the level sensitivity, see Eq.~\eqref{rth} in the main text, means that $R \ge {\rm const}/(L^{m+1} \Delta \epsilon)$. Since the level spacing, $\Delta \epsilon$, decays exponentially with $L$, the power-law term, $L^{m+1}$, may introduce only logarithmic corrections, $\sim [\log(\Delta \epsilon)]^{m+1}$, to the latter dependence. Consequently, Eq.~\eqref{assums} is sufficient for the level sensitivity to exhibit a significant decrease upon introducing IBP from $R\sim 1/\Delta \epsilon$ (in the integrable case) to $R\sim 1/\sqrt{\Delta \epsilon}$ (in the chaotic system). 
We stress that the bound on the right-hand side of  Eq.~\eqref{bounds} is not tight, however it is sufficient for the purpose of the present discussion.
}

\subsection{Nonzero magnetization sectors}

\zm{In the main text, we have discussed results obtained for $S^z_{tot}=0$. In the case of integrable chain, we have found substantial differences between the level sensitivities obtained in the canonical ensemble ($R_{CE}$) and the grand canonical ensemble ($R_{GCE}$). This indicates that $R_{CE}$ for $S^z_{tot}=0$ must be very different from $R_{CE}$ in sectors with other magnetization, $m=S^z_{tot}/L$. However, one expects that both ensembles become equivalent at least in the chaotic regime.} 

\begin{figure}[hptb]
    \centering
    \includegraphics[width=1.0\linewidth]{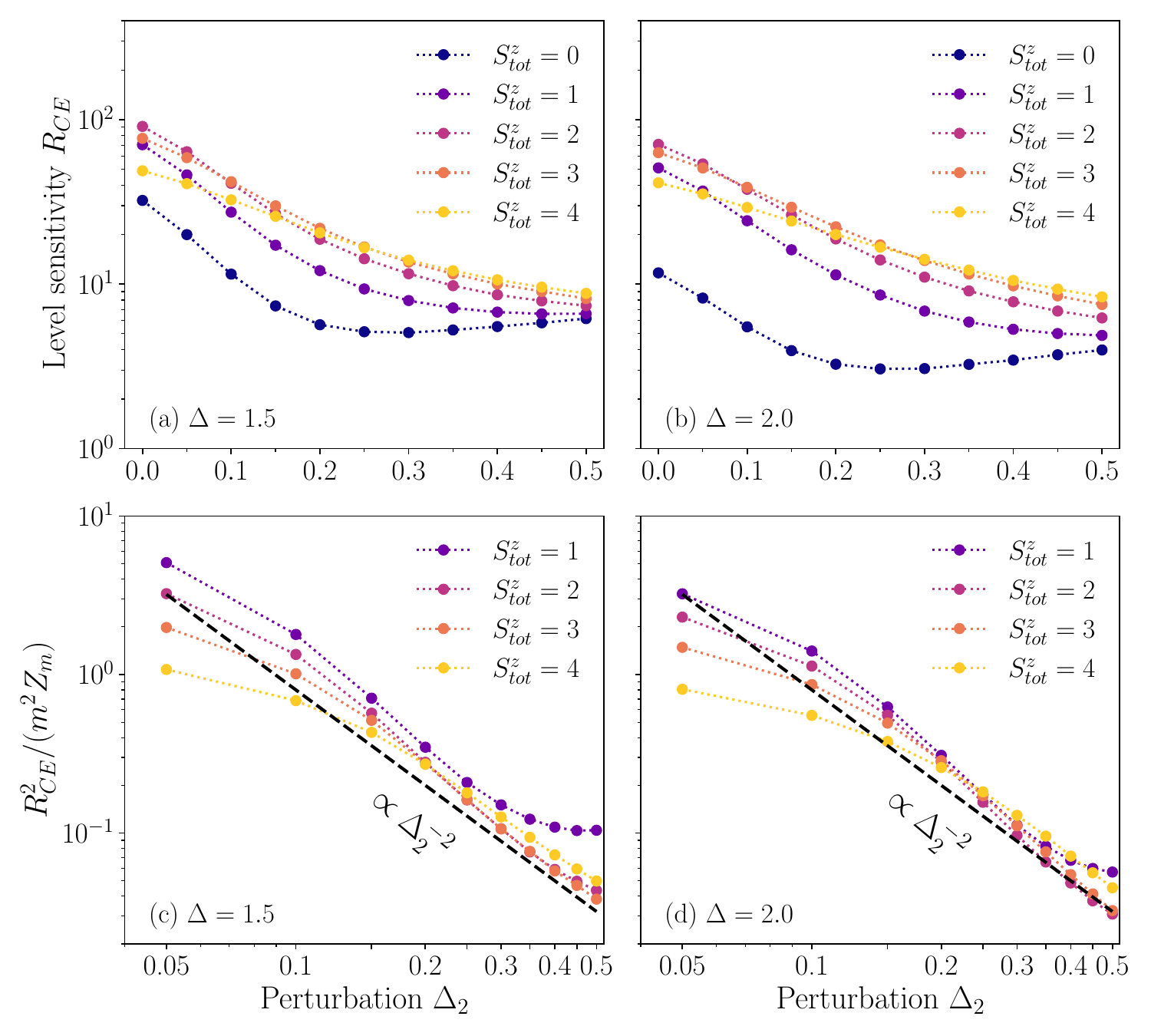}
    \caption{Level sensitivity $R_{CE}$ vs. perturbation strength $\Delta_2$ within magnetization sectors $S^z_{tot} = 0,1,2,3,4$, calculated via full ED in system with $L=20$, for (a) $\Delta=1.5$ and (b) $\Delta=2.0$. In panels (c) and (d) we show square of level sensitivity rescaled by the Hilbert space dimension and $m^2$, for $\Delta=1.5$ and $\Delta=2.0$ respectively. Dashed black line indicates a power-law dependence $\propto (\Delta_2)^{-2}$.}
    \label{sm_nw6}
\end{figure}
\zm{
In order to confirm both these expectations, we study the XXZ model [Eq.~(\ref{xxz}) in the main text] with perturbation  $g=\Delta_2$ and $H^\prime = \sum_l S^z_{l+2} S^z_l$. Figures  \ref{sm_nw6}(a) and  \ref{sm_nw6}(b) show
$R_{CE}$ as a functions of $\Delta_2$ for $L=20$ and $S^z_{tot}=0,1,2,3,4$. Here, $S^z_{tot}=1$ means that we applied one spin flip with respect to $S^z_{tot}=0$. One observes strong 
$S^z_{tot}$-dependence for $\Delta_2=0$ that diminishes when the integrability-breaking perturbation increases. For $\Delta > \Delta^*$, i.e. in the chaotic regime, $R_{CE}$ appears to change smoothly with magnetization. This smooth dependence is consistent with the equivalence of CE and GCE. 
}

\zm{Finally, we demonstrate for the chaotic regime that the magnetization-dependence of $R_{CE}$ 
is consistent with $S^z_{tot}$-dependence of the spin conductivity. The latter consistency is set by the RMT-relation [Eq.~(\ref{rth1}) in the main text]. Spin conductivity, $\sigma$, has been studied numerically in Ref. \cite{prelovsek22}, see Appendix A therein, for sectors with various magnetization,  $m=S^z_{tot}/L$. The following relation has been established
\begin{eqnarray}
\sigma(m) & \simeq & \sigma(m=0)+{\rm const} \frac{m^2}{(\Delta_2)^2} \\  & \simeq &{\rm const} \frac{m^2}{(\Delta_2)^2}, \label{s_sig}
\end{eqnarray}
and in Eq.~(\ref{s_sig}) we assume that $m/\Delta_2$ is sufficiently large. Following Eq.~(\ref{rth1}) in the main text, one expects that $R^2_{CE} \Delta \epsilon$ should exhibit the same dependence on $m$ and $\Delta_2$. As a rough estimate of the level spacing we take $\Delta \epsilon\propto 1/Z_m$, where $Z_m$ is the dimension of the subspace with magnetization $m$.  Figures  \ref{sm_nw6}(c) and  \ref{sm_nw6}(d) show $R^2_{CE}/(m^2 Z_m)$
for $S^z_{tot}=1,2,3,4$. In the chaotic regime one observes nearly overlapping curves which decay as $1/(\Delta_2)^2$.  The deviations from this dependence show up, as expected, for small $m/\Delta_2$.  
}

\end{document}